\documentclass{PoS}
\usepackage{epsfig}

\title{Chiral perturbation theory, $K\rightarrow \pi \pi$ decays and 2+1 flavor domain wall QCD }

\ShortTitle{Chiral perturbation theory, $K\rightarrow \pi \pi$ decays and 2+1 flavor domain wall QCD }

\author{Shu Li\\
        Columbia University, USA\\
        E-mail: \email{lishu@phys.columbia.edu}}

\author{\speaker{Norman Christ}
         \thanks{This work was partially supported by DOE grant 
         \#DE-FG02-92ER40699.  The authors thank RIKEN, BNL, the 
         US DOE, PPARC and the University of Edinburgh for providing
         the facilities essential for this work.}\\
        Columbia University, USA\\
        E-mail: \email{nhc@phys.columbia.edu}}

\author{RBC and UKQCD collaborations}

\abstract{We present a calculation of the low energy 
constants describing the real and imaginary parts of the 
$K \rightarrow \pi \pi$ decay amplitudes $A_0$ and $A_2$.  
Leading and next leading order chiral perturbation theory 
is used and its applicability assessed.  A combination of 
statistical and systematic errors limits the precision of 
the results. The apparent limitations of chiral perturbation 
theory raise doubts about the accuracy of a possible 
extrapolation to physical $K\rightarrow \pi \pi$ 
kinematics.}

\FullConference{The XXVI International Symposium on Lattice Field Theory \\
		 July 14 - 19, 2008\\
		 Williamsburg, Virginia, USA}

\begin{document}

Quantitative understanding of the two pion decay of the
K meson has been an outstanding problem in particle physics
for the past fifty years.  While much progress has been made
in explaining the large size of the $\Delta I = 1/2$
amplitude relative to that with $\Delta I = 3/2$, a precise
prediction of this ``$\Delta I = 1/2$ rule'', has not been
achieved.  Of even greater interest is determining if the direct 
CP violation observed in $K_L \rightarrow \pi\pi$ decay can be 
computed from the single CP violating phase present in the CKM 
matrix or whether new physics is required.

Lattice QCD offers the promise of a first-principles calculation
of these quantities.  However, two serious difficulties must be 
overcome.  The first arises from the complexity of the low energy 
weak Lagrangian whose matrix elements determine the needed decay 
amplitudes.  There are seven independent four-quark operators.  
In an environment with accurate flavor and chiral symmetries, 
these seven operators mix in groups of one, two and four and also 
with operators of lower dimension.  Without flavor and chiral 
symmetry, a much larger class of operators must be studied.

The second difficulty comes from the two-pion final state.  The 
two-pion state isolated at asymptotically large time in a Euclidean 
lattice QCD calculation will be the lowest energy state with the
given quantum numbers.  Thus, the $I=2$ state will be two pions
at nearly zero relative momentum, not the physical value of 
$p \approx 205$ MeV.   For the $I=0$ case, the lowest energy state 
will be the vacuum.

The first difficulty appears to now be largely overcome.  By using
a chiral fermion formulation, {\it e.g.} domain wall fermions, and
the Rome-Southampton RI/MOM normalization procedure it is possible
to isolate and properly normalize the seven relevant weak operators,
accurately including the effects of operator mixing.  This was 
developed and shown to be practical in our earlier quenched 
calculation seven years ago~\cite{Blum:2001xb}, a reference defining
the conventions used here.

In this talk, we attempt to avoid the problems created by the
two-pion final state by using chiral perturbation theory (ChPT)
to relate the $K \rightarrow \pi\pi$ amplitude of interest to 
$K \rightarrow \pi$ and $K \rightarrow |0\rangle$ matrix elements 
which are much easier to evaluate using lattice methods
\cite{Bernard:1985wf}.  This method was used in Ref.~\cite{Blum:2001xb}.  
However, the series of quark masses used in that calculation were 
too large in size and too few in number to provide a test of chiral 
perturbation theory.

The most apparent difficulty in that earlier calculation is the
use of the quenched approximation.  As discovered by Golterman
and Pallente~\cite{Golterman:2001qj}, quenched chiral perturbation 
theory for the (8,1) amplitudes reveals significant, unphysical 
logarithms not present in full QCD calculation and subsequent 
numerical studies of these terms suggested that they could be 
large~\cite{Aubin:2006vt}.

The present calculation, reported here, removes this problem by 
using the 2+1 flavor lattice ensembles of the RBC and UKQCD
collaborations.   By including both unitary and partially quenched
quark masses and working with somewhat lighter masses, the present
calculation also provides information about the validity of 
chiral perturbation theory.  In addition, the recent 2+1 flavor, 
partially quenched chiral perturbation theory (PQChPT) calculation 
of Aubin, Laiho, Li and Lin~\cite{Aubin:2008vh} provides the theoretical 
formulae necessary to extract low energy constants from our results 
and to explore the validity of this approach.

As we will see, $SU(3)\times SU(3)$ chiral perturbation theory 
describes our results poorly.  This same conclusion was reached 
in a study of meson masses, decay constants and neutral kaon mixing 
using these same 2+1 flavor ensembles \cite{Allton:2008pn}.  
However, in the present case the use of $SU(3)\times SU(3)$ chiral 
perturbation theory is required by our strategy for calculating 
two-pion matrix elements, making the final results of the calculation
presented here highly uncertain.

\section{Description of the calculation}

This calculation is based on the $24^3 \times 64$ RBC/UKQCD 2+1 
flavor ensembles.  These have an inverse lattice spacing of 1.73 
GeV and a linear extent of 2.7 fm.  Two ensembles are examined: one 
with a light quark mass $m_l=0.005$ corresponding to a pion mass 
of 331 MeV and the second with $m_l=.01$ and a pion mass of 415 MeV.  
Both use 0.04 for the strange mass, approximately 15\% larger than 
that of the physical strange quark.  The fifth-dimension extent of 
the domain fermion lattice is $L_s=16$ and the resulting residual 
mass 0.00315(2).

\begin{figure}[ht]
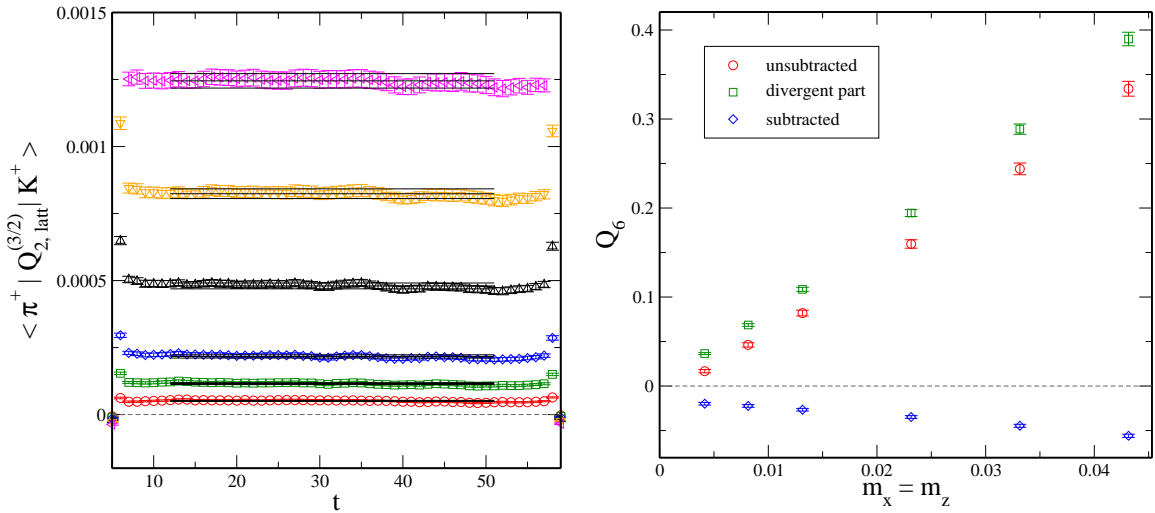

    \hfill
    \begin{minipage}{0.45\textwidth}
        \hspace{-1.0cm}
        \epsfig{file=O2_plat_deg.eps,width=1.1\linewidth,angle=0} 
    \end{minipage}
    \hfill
    \begin{minipage}{0.45\textwidth}
        \hspace{-0.8cm}
        \epsfig{file=O6PQS_vacsub_subtd_mres.eps,
                       width=1.1\linewidth,angle=0}
    \end{minipage}
\caption{The left panel shows the dependence of $\pi-Q_2-K$ amplitude
on the location $t$ of the $Q_2$ operator at which the four quark
propagators coming from the sources at $t=5$ and 59 are combined.  
From the top the curves describe the degenerate light quark masses 
0.001, 0.005, 0.01, 0.02, 0.03, 0.04.  The right panel shows the 
matrix elements of $Q_6$ (circles), the subtraction term (squares) 
and their difference (diamonds).}
\label{fig:plateau-subtraction}
\hfill
\vspace{-0.4cm}
\end{figure}

The weak matrix elements are computed using a Coulomb gauge-fixed 
wall source for the quarks making up the pion located at time slice 
$t=5$ while the quarks in the $K$ meson are created by a similar wall 
source located at $t=59$.  Interference from quarks which travel 
through the time boundary at $t=63-0$ is suppressed by ``doubling'' 
the lattice, achieved by determining the quark propagators twice,
first using periodic and second using anti-periodic boundary 
conditions in the time.  The average of these propagators will 
have a source at $t=5$ or 59 and no image at $t=69$ or $t=-6$, 
respectively.  As shown in Fig.~\ref{fig:plateau-subtraction}, 
even with this large separation of 64 between the two sources, 
the 3-point function is quite accurately determined with some 
error reduction coming from the average over the long plateau.

A particularly delicate part of the calculation is associated with
the mixing of the (8,1) operators with the lower dimensional operators
$\overline{s}d$ and $\overline{s}\gamma^5 d$.  While these quadratically
divergent contributions, ({\it i.e.} $\sim 1/a^2$) vanish when the 
weak operator carries no four-momentum, these operators do contribute
when chiral perturbation theory is used.  However, just as in our
earlier quenched work, with proper care this subtraction, determined 
by the ChPT expression, can be done to better than 10\% accuracy.  The 
right panel of figure~\ref{fig:plateau-subtraction} shows the two 
amplitudes which must be subtracted and their well-resolved difference.

\section{Chiral extrapolation}

We first discuss the $\Delta I = 3/2$ LEC $\alpha_{27}$ which 
can be determined from the chiral limit of the matrix element 
$\langle \pi|Q^{(27,1)}|K\rangle$.  This operator makes up
the $\Delta I = 3/2$ part of $Q_1$, $Q_2$, $Q_9$ and $Q_{10}$
and with $K-\overline{K}$ external states determines $B_K$.  
Following the  ChPT studies in Ref~\cite{Allton:2008pn}, we 
limit the range of input masses in our chiral fits to those 
whose average is 0.01 or less in lattice units.  We were 
unable to obtain a sensible, NLO chiral fit to this quantity.  
While it was possible to describe our data by the NLO chiral 
formula, the large chiral log with coefficient of -34/3 was 
sufficiently inconsistent with our data and the other LEC's
sufficiently unconstrained, that the resulting fits gave a 
leading order term contributing 3\% to the total with the 
next leading order terms providing 97\%.  (When defined in 
the same fashion the coefficient of the NLO logarithm in 
$m_\pi^2$ is 2/3.)

\begin{figure}[ht]
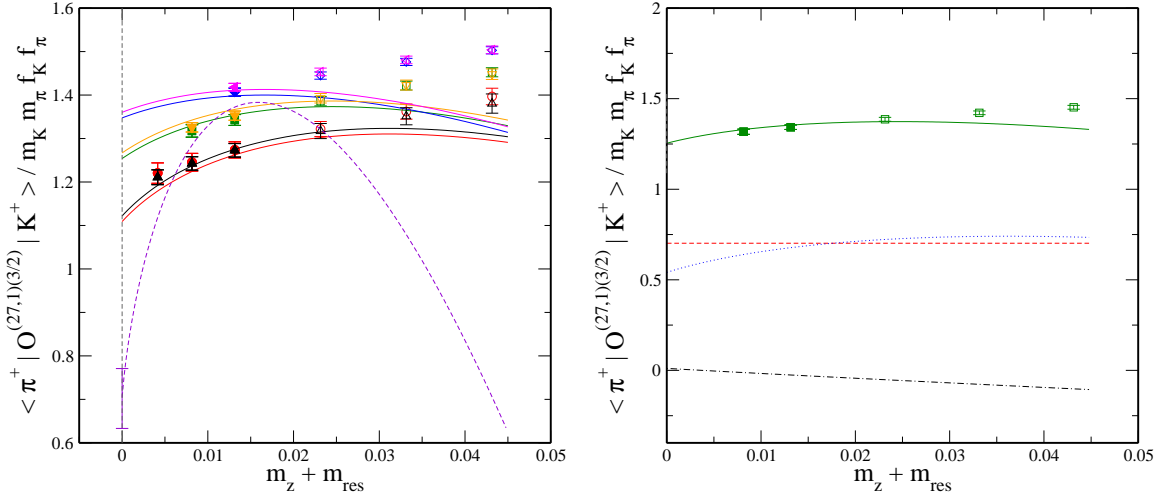

    \vspace{-0.1cm}
    \hfill
    \begin{minipage}{0.45\textwidth}
        \hspace{-1.0cm}
        \epsfig{file=i32kpi_i32only.eps,width=1.1\linewidth,angle=0} 
    \end{minipage}
    \hfill
    \begin{minipage}{0.45\textwidth}
        \hspace{-0.8cm}
        \epsfig{file=i32kpi_i32only_term_div.eps,
                       width=1.1\linewidth,angle=0}
    \end{minipage}
\caption{The left panel shows the NLO ChPT fit to the ratio
$\langle \pi|Q^{(27,1)}|K\rangle/(f_Kf_\pi m_K^2 m_\pi^2)$.  
The three bands of points from bottom to top correspond to
light quark masses of 0.001, 0.005 and 0.01.  In each band of
points the lower have a light sea quark of mass 0.005 and upper,
0.01.  The solid lines are the chiral fits and the dotted line
a unitary extrapolation (based on those fits) to the chiral 
limit.  In the right-hand panel contributions of the individual 
terms are broken out and the uncomfortably large size of the NLO
terms shown.  The solid curve passing through the data points is 
the sum of all terms.  The next two moving downward are the 
leading order term and the NLO logarithms.  The bottom curve is 
the NLO analytic terms.}
\label{fig:3-2_Ch_fit}
\hfill
\vspace{-0.6cm}
\end{figure}

In order to sidestep this difficulty, we next fit the ratio 
$\langle \pi|Q^{(27,1)}|K\rangle/(f_K f_\pi m_K^2 m_\pi^2)$, 
similar to the amplitude giving $B_K$, at NLO.  Taking this 
ratio reduces the statistical errors in our computed quantities 
and reduces the coefficient of the chiral logarithm to -6.  The 
NLO fit to this ratio is more satisfactory and is shown in 
Fig.~\ref{fig:3-2_Ch_fit}.  While the left-hand panel shows a 
reasonable chiral fit to this ratio, the right-hand panel shows 
the contribution of the various terms in the chiral expansion 
and reveals that the NLO correction is of the same size as the 
LO term.  In order to test the robustness of our use of this 
{\it ad hoc} ratio, we also divided by a second factor of 
$f_K f_\pi$ and found similar results for the corresponding NLO
ChPT fit.  The low energy constants obtained in the fit to the 
first ratio, for each of the $\Delta I = 3/2$ operators 
$Q^{(27,1)}$, $Q_7$ and $Q_8$ are shown in Table~\ref{tab:LECs}.  
The statistical error obtained by a standard jackknife analysis 
is shown in the left bracket.  Our estimate of a systematic 
error is given in the right bracket.  

\begin{table}
\centering
\begin{tabular}{ccc}
\hline
\hline
$ Q_i $ & $\alpha_{i,{\rm ren}}^{(1/2)}$ & $\alpha_{i,{\rm ren}}^{(3/2)}$ \\
\hline
1 & $ -6.6(15)(66) \times 10 ^ { -5 } $ & $ -2.48(24)(39) \times 10 ^ { -6 } $\\
2 & $ 9.9(21)(99)   \times 10 ^ { -5 } $ & $ -2.47(24)(39) \times 10 ^ { -6 } $\\
3 & $ -0.8(31)(21) \times 10 ^ { -5 } $ & 0.0\\
4 & $ 1.62(44)(162) \times 10 ^ { -4 } $ & 0.0\\
5 & $ -1.52(29)(152) \times 10 ^ { -4 } $ & 0.0\\
6 & $ -4.1(7)(41)    \times 10 ^ { -4 } $ & 0.0\\
7 & $ -1.11(17)(18) \times 10 ^ { -5 } $ & $ -5.53(85)(91) \times 10 ^ { -6 } $\\
8 & $ -4.92(72)(75) \times 10 ^ { -5 } $ & $ -2.46(37)(37) \times 10 ^ { -5 } $\\
9 & $ -9.8(20)(98) \times 10 ^ { -5 } $ & $ -3.72(37)(59) \times 10 ^ { -6 } $\\
10 & $ 6.8(15)(68) \times 10 ^ { -5 } $ & $ -3.69(37)(59) \times 10 ^ { -6 } $\\
\hline
\hline
\end{tabular}
\caption{The values for the low energy constants which describe 
the ten operators $Q_1,...,Q_{10}$ at leading order in ChPT.  These 
correspond to operators renormalized in the $\overline{\mbox{MS}}$ 
scheme at the scale $\mu = 2.15$ GeV.  The LEC's for $Q_1,...Q_6$, 
$Q_9$ and $Q_{10}$ are expressed in (GeV)$^4$ while those for $Q_7$ 
and $Q_8$ are given in (GeV)$^6$.}
\label{tab:LECs}
\end{table}

Given the failure of simple NLO PQChPT to directly describe the 
matrix element $\langle \pi|Q^{(27,1)}|K\rangle$, assessing the 
systematic error in a chiral extrapolation of this quantity is 
difficult.  The systematic errors presented for the three 
$\Delta I = 3/2$ LECs in Table~\ref{tab:LECs} were obtained by 
computing the difference between the results from the NLO chiral 
fits to the two ratios described above and then doubling the 
result in an attempt to account for the uncertainties in this 
{\it ad hoc} procedure and the uncomfortably large relative size 
of the NLO terms.

\begin{figure}[ht]
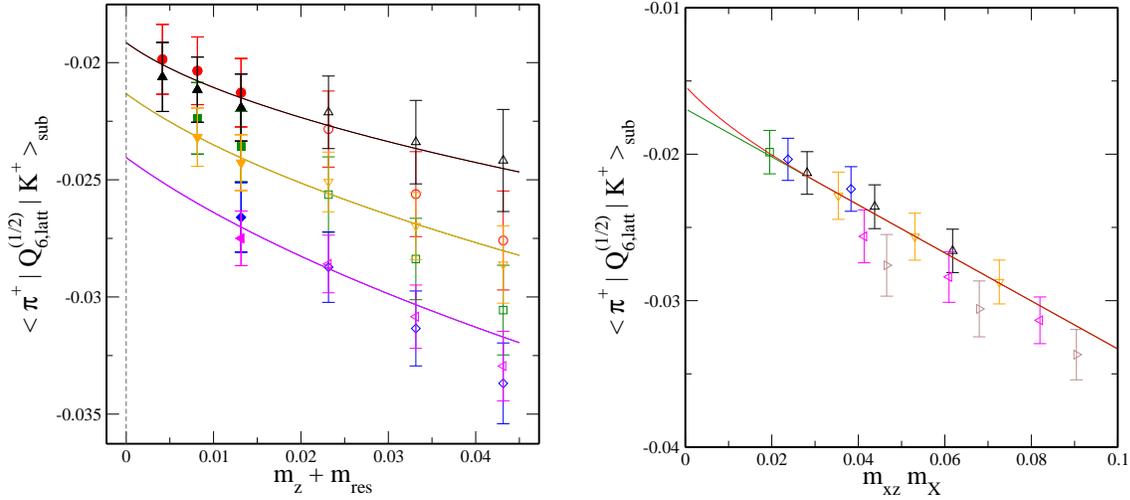

    \vspace{-0.3cm}
    \hfill
    \begin{minipage}{0.45\textwidth}
        \hspace{-1.0cm}
        \epsfig{file=O6PQS_kpi_LOkpi_withmres.eps,
                       width=1.05\linewidth,angle=0} 
    \end{minipage}
    \hfill
    \begin{minipage}{0.45\textwidth}
        \hspace{-0.8cm}
        \epsfig{file=O6PQS_syserr.eps,
                       width=1.05\linewidth,angle=0}
    \end{minipage}
\caption{The left panel shows the LO ChPT fit to the matrix
element $\langle \pi|Q_6|K\rangle$.  The ordering of the data
is similar to that in left panel of Fig. 2 except that the 
light valence quark mass now increases from top to bottom.  The 
data is replotted in the right panel as a function of product 
of the partially quenched pion and kaon masses.  The second, 
upward bending curve estimates the effects of the omitted chiral 
logarithms.}
\label{fig:1-2_Ch_fit}
\hfill
\vspace{-0.4cm}
\end{figure}

Next we describe the extraction of the $\Delta I = 1/2$ low 
energy constants.   Here difficulties arose because of the 
large number of LECs which appear in the NLO expression for an 
(8,1) operator.  Our twelve PQ data points with average input 
quark masses at or below the 0.01 upper limit proved insufficient
to give a stable fit to the required eight LO and NLO LECs.  As 
a result we could perform only the LO fit shown in left panel
of Fig~\ref{fig:1-2_Ch_fit} for the case of $Q_6$.  As can 
be seen, the fit nicely describes both the data included in the 
fit and that at larger masses as well.  However, the omitted 
chiral logarithms can substantially alter the resulting LEC (the 
slope of the correct curve, including the logarithms, at $m=0$). 
This difficulty is shown in the right panel of Fig~\ref{fig:1-2_Ch_fit}
where the second, upward bending curve at $m=0$ is obtained by
adding an $m_\pi^4 \ln (m_\pi^2/\Lambda^2)$ term with the correct 
coefficient for the case of vanishing degenerate light quark masses
but fixed strange quark mass.  The constant $\Lambda$ as well as 
the two parameters in the altered linear terms are matched to the 
LO linear fit in value, slope and curvature at our lightest 
dynamical quark mass $(m_l = 0.005)$.  As can be seen in the 
figure, this increases the slope at zero by a factor of two.  

The resulting LEC for the $\Delta I = 1/2$ operators are also
shown in Table~\ref{tab:LECs}.  The 100\% systematic errors
arise from our inability to constrain the place at which the
simple linear behavior seen in our results is replaced by the
proper, non-linear, logarithmic behavior required by ChPT.  The 
factor of two increase in $\alpha_6$ discussed above is typical 
for these amplitudes.  Since a larger matching point than 0.005 
is consistent with our data and leads to even larger effects, 
we do not believe that 100\% is an obvious over estimate of this
uncertainty.

\begin{figure}[ht]
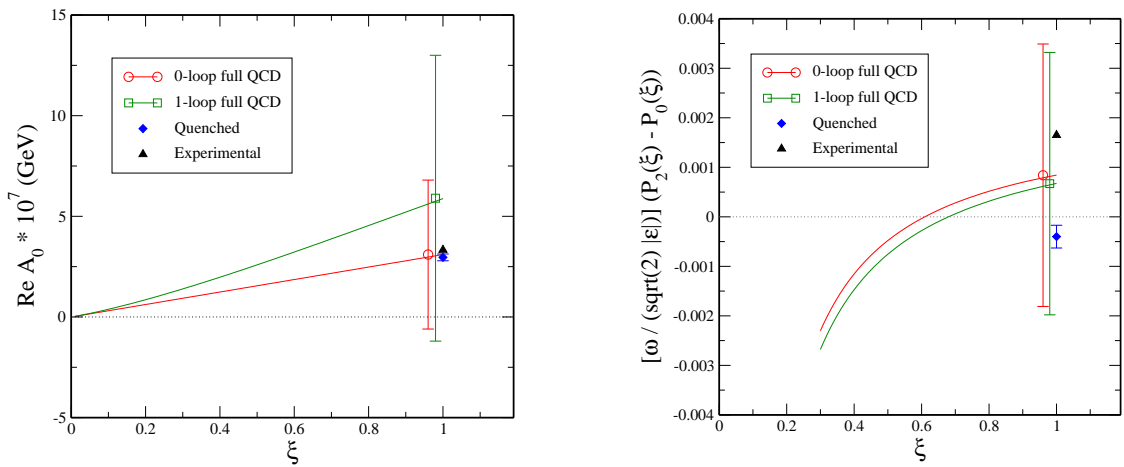

    \vspace{-0.3cm}
    \hfill
    \begin{minipage}{0.45\textwidth}
        \hspace{-1.0cm}
        \epsfig{file=ReA0_xi.eps,width=1.0\linewidth,angle=0} 
    \end{minipage}
    \hfill
    \begin{minipage}{0.45\textwidth}
        \hspace{-0.2cm}
        \epsfig{file=ReEE_xi.eps,
                       width=0.95\linewidth,angle=0}
    \end{minipage}
\caption{The left panel shows the real part of $A_0$ as a function 
the factor $\zeta$ by which both $m_\pi^2$ and $m_K^2$ have been 
scaled to allow a connection to be made with the region where
ChPT should become accurate.  The lower curve is a LO ChPT 
extrapolation while the upper curve includes the NLO chiral 
logarithm.  The right panel shows similar LO and NLO logarithmic 
extrapolations for the quantity equal to Re$(\epsilon^\prime/\epsilon)$ 
when $\zeta=1$.  The triangles show the experimental values while 
the diamonds the quenched results of Ref~\cite{Blum:2001xb}.}
\label{fig:Ch_extrap}
\hfill
\vspace{-0.4cm}
\end{figure}

Finally we attempt to combine the LECs determined above to
determine the real and imaginary parts of the $I=0$ and 2, 
$K \rightarrow \pi\pi$ decay amplitudes $A_0$ and $A_2$.
Such a determination is made highly uncertain by the failure
of chiral perturbation theory to describe physics for quark
masses in the region of the strange quark as discussed above
and in Ref~\cite{Allton:2008pn}.  This leads to the large 
uncertainties in the necessary LECs in Table~\ref{tab:LECs} 
and makes the extrapolation from those LECs to the physical
$K \rightarrow \pi\pi$ amplitudes unreliable.  Further, the 
needed 2+1 flavor NLO ChPT formulae are not available nor 
have we determined full the set of LEC's needed for such a 
NLO extrapolation.  The resulting values for the physical
quantities of interest are given in Table~\ref{tab:results}.
The large systematic errors are determined by adding in 
quadrature the propagated systematic errors from the LO LECs 
and the difference between a simple LO extrapolation and a 
NLO extrapolation in which only the effects of the logarithms
are included, using the chiral scale $\Lambda = 1$ GeV but
with no NLO analytic terms.  Fits with and without the 
logarithm terms are compared in Figure~\ref{fig:Ch_extrap}.

\begin{table}
\centering
\begin{tabular}{cccc}
 \hline
 \hline
 Quantity & This analysis & Quenched~\cite{Blum:2001xb} & Experiment \\
 \hline
 ${\rm Re}A_0$ (GeV) & $4.5(11)(53)\times 10^{-7}$ & $2.96(17) \times 10^{-7}$ & $3.33 \times10^{-7}$\\
 ${\rm Re}A_2$ (GeV) & $8.57(99)(300)\times 10^{-9}$ & $1.172(53) \times 10^{-8}$ & $1.50\times 10^{-8}$\\
 ${\rm Im}A_0$ (GeV) & $-6.5(18)(77)\times10^{-11}$ & $-2.35(40)\times10^{-11}$ & \\
 ${\rm Im}A_2$ (GeV) & $-7.9(16)(39)\times10^{-13}$ & $-1.264(72)\times10^{-12}$ & \\
 $1/\omega$	     & $50(13)(62)$	& $25.3(1.8)$	& 22.2 \\
 ${\rm Re}(\epsilon'/\epsilon)$ & $7.6(68)(256)\times10^{-4}$ & $-4.0(2.3)\times10^{-4}$ & $1.65\times10^{-3}$ \\
 \hline
 \hline
\end{tabular}
\caption{Values for physical quantities describing 
$K \rightarrow \pi \pi$ decay.  Here $\omega = \mbox{Re}A_2/\mbox{Re}A_0$.
Note, only statistical errors are given for the quenched results.}
\label{tab:results}
\end{table}

We have presented preliminary first results for $K \rightarrow \pi\pi$ 
decays determined by lattice calculation and chiral perturbation 
theory.  These have been obtained in a theory with accurate 
chiral symmetry, with a proper treatment in full QCD of the 
light and strange quarks and a complete renormalization, 
including the effects of operator mixing, of the seven 
independent four-quark operators needed to describe this decay.  
The errors on the $\Delta I = 3/2$ LECs could be reduced by 
extending our calculation to smaller quark masses while those 
for $\Delta I = 1/2$ could be decreased if more quark mass 
combinations were studied.  Likewise some uncertainties related 
to the extrapolation to the physical $K \rightarrow \pi\pi$ 
amplitudes could be reduced if the full NLO ChPT formulae 
were available.  However, these steps would not reduce the
large, dominate uncertainties associated with the use of 
ChPT at the scale of the kaon as is required by this 
treatment the $\pi$-$\pi$ final state.  

We conclude that useful results for these important quantities
require the direct study of $\pi$-$\pi$ final states. The RBC 
and UKQCD collaborations are now actively pursing calculations 
with larger volumes and lattice spacings~\cite{Renfrew:2008xx} 
and lighter pions, working toward this goal.

\bibliographystyle{JHEP}
\bibliography{lat08}

\providecommand{\href}[2]{#2}\begingroup\raggedright\begin{thebibliography}{1}

\bibitem{Blum:2001xb}
{\bf RBC} Collaboration, T.~Blum {\em et~al.}, {\it Kaon matrix elements and
  cp-violation from quenched lattice qcd. i: The 3-flavor case},  {\em Phys.
  Rev.} {\bf D68} (2003) 114506
  [\href{http://arXiv.org/abs/hep-lat/0110075}{{\tt hep-lat/0110075}}].

\bibitem{Bernard:1985wf}
C.~Bernard, T.~Draper, A.~Soni, H.~D. Politzer and M.~B. Wise, {\it Application
  of chiral perturbation theory to k $\to$ 2 pi decays},  {\em Phys. Rev.} {\bf
  D32} (1985) 2343.

\bibitem{Golterman:2001qj}
M.~Golterman and E.~Pallante, {\it Effects of quenching and partial quenching
  on penguin matrix elements},  {\em JHEP} {\bf 10} (2001) 037
  [\href{http://arXiv.org/abs/hep-lat/0108010}{{\tt hep-lat/0108010}}].

\bibitem{Aubin:2006vt}
C.~Aubin {\em et~al.}, {\it {Systematic effects of the quenched approximation
  on the strong penguin contribution to epsilon'/epsilon}},  {\em Phys. Rev.}
  {\bf D74} (2006) 034510 [\href{http://arXiv.org/abs/hep-lat/0603025}{{\tt
  arXiv:hep-lat/0603025}}].

\bibitem{Aubin:2008vh}
C.~Aubin, J.~Laiho, S.~Li and M.~F. Lin, {\it {K to pi and K to 0 in 2+1 Flavor
  Partially Quenched Chiral Perturbation Theory}},
  \href{http://arXiv.org/abs/0808.3264}{{\tt arXiv:0808.3264 [hep-lat]}}.

\bibitem{Allton:2008pn}
C.~Allton {\em et~al.}, {\it {Physical Results from 2+1 Flavor Domain Wall QCD
  and SU(2) Chiral Perturbation Theory}},
  \href{http://arXiv.org/abs/0804.0473}{{\tt arXiv:0804.0473 [hep-lat]}}.

\bibitem{Renfrew:2008xx}
D.~Renfrew {\em et~al.}, {\it Controlling residual chiral symmetry breaking in
  domain wall fermion simulations},  {\em PoS} {\bf LAT2008} (2008) 048.

\end{thebibliography}\endgroup


\end{document}